\documentclass[12pt,preprint]{aastex}

\begin{document}

\title{Weak Lensing Mass Reconstruction: Flexion vs Shear}
\author{S.~Pires}
\affil{Laboratoire AIM, CEA/DSM-CNRS-Universite Paris Diderot, IRFU/SEDI-SAP, Service d'Astrophysique, \\CEA Saclay, Orme des Merisiers, 91191 Gif-sur-Yvette, France}
\and
\author{A.~Amara}
\affil{Department of Physics, ETH Zurich, Wolfgang-Pauli-Strasse 16, CH-8093 Zurich, Switzerland}

\begin{abstract}
Weak gravitational lensing has proven to be a powerful tool to map directly the distribution of dark matter in the Universe. The technique, currently used, relies on the accurate measurement of the gravitational shear that corresponds to the first-order distortion of the background galaxy images. More recently, a new technique has been introduced that relies on the accurate measurement of the gravitational flexion that corresponds to the second-order distortion of the background galaxy images. This technique should probe structures on smaller scales than that  of a shear analysis. The goal of this paper is to compare the ability of shear and flexion to reconstruct the dark matter distribution by taking into account the dispersion in shear and flexion measurements.  
Our results show that the flexion is less sensitive than shear for constructing the convergence maps on scales that are physically feasible for mapping, meaning that flexion alone should not be used to do convergence map reconstruction, even on small scales.
\end{abstract}


\section{Introduction}
Weak gravitational lensing is a powerful tool for mapping the distribution of dark matter since it measures the matter distribution directly without the need to make assumptions about the way that light traces mass. Most approaches focus on shear, which is the first-order distortion of the background galaxy images caused by the bending of light from Large Scale Structure (LSS). Several methods have been developed to reconstruct the projected mass distribution from the observed shear field \citep[e.g.][]{astro:kaiser93,wlens:seitz98,wlens:bridle98,wlens:marshall02,wlens:starck06}.

Recently, weak lensing techniques have been extended to include higher-order distortions of background galaxies to improve the constraints on the mass distribution on small scales. The measurement of the second-order distortion of the background galaxy images by means of the galaxy octopole moments was introduced by \cite{flexion:goldberg02}. This second-order in image distortions corresponding to a third-order effect in gravitational potential \citep{flexion:bacon06} is responsible for the weakly skewed and arc-like appearance of lensed galaxies and is expected  to probe variations of the gravitational potential field on smaller scales than those accessible by shear analysis alone. Despite the fact that measurement of the octopole moments is more complex, their intrinsic dispersion due to the random shapes of galaxies is expected to be much smaller than the intrinsic ellipticity dispersion. In \cite{flexion:goldberg05}, the method has been further developed by using the shapelet formalism to estimate second-order lensing effect. At the same time, a related approach using the galaxy sextupole moments has also been explored \citep{flexion:irwin03,flexion:irwin05,flexion:irwin06}.  In \cite{flexion:okura07}, the authors suggest a new method called HOLICs to measure the second-order lensing effect based on the measurement of the octopole and higer-order moments.  In \cite{flexion:goldberg05}, the second-order lensing effect was detected for the first time and the term ``flexion" has been adopted to describe it. The formalism to reconstruct the projected mass distribution from the flexion measurements was introduced by \cite{flexion:bacon06} for the first time. 

This paper is structured as follows. In \S 2, we review the basis of the weak gravitational lensing and the flexion formalism. We then proceed with an introduction to the mass inversion problem from shear and flexion measurements. In \S 3, a comparison between shear and flexion is conducted in order to compare their ability to reconstruct the convergence map in the presence of noise.  
In \S 4, we have a discussion about published results on mass map reconstruction from flexion.
In \S 5, we conclude in the implications of our results on future flexion studies.

\section{Review of weak gravitational lensing formalism}
\subsection{Shear formalism}

\subsubsection{The second-order of the gravitational lensing potential $\psi$}

Shear $\gamma_i({\mathbf \theta})$ with $i=1,2$ is measured from the shapes of galaxies at positions ${\mathbf\theta}$ in an image. The shear field $\gamma_i({\mathbf \theta})$ can be written in terms of the lensing potential $\psi({\mathbf
\theta})$ as \citep[see e.g.][]{bartelmann99}:
\begin{eqnarray}
\label{eq:gamma_psi} 
\gamma_1 & = & \frac{1}{2}\left( \partial_1^2 -
\partial_2^2 \right) \psi, \nonumber \\
\gamma_2 & = & \partial_1
\partial_2 \psi,
\end{eqnarray}
where the partial derivatives $\partial_i$ are with respect to $\theta_i$.  The convergence $\kappa({\mathbf \theta})$ can also be written in terms of the lensing potential as:
\begin{equation}
\label{eq:kappa_psi}
\kappa =  \frac{1}{2}\left(\partial_1^2 + \partial_2^2 \right) \psi.
\end{equation}
The convergence $\kappa$ corresponds to the projected (normalized) mass distribution. \\

\begin{figure*}
\centerline{
\hbox{
\includegraphics[width=5.6cm,height=6.5cm]{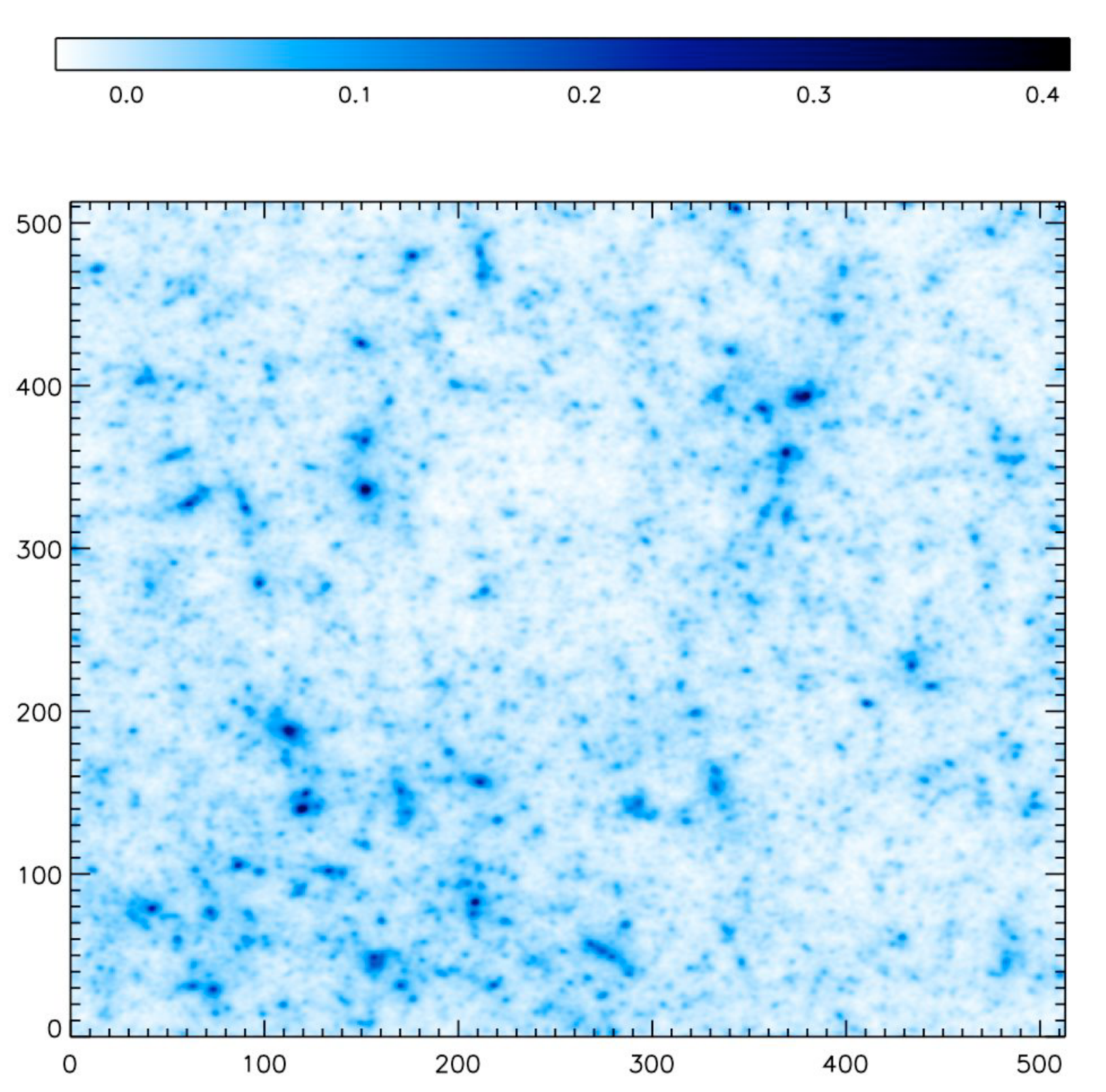}
\hspace{0.2cm}
\includegraphics[width=5.6cm, height=6.5cm]{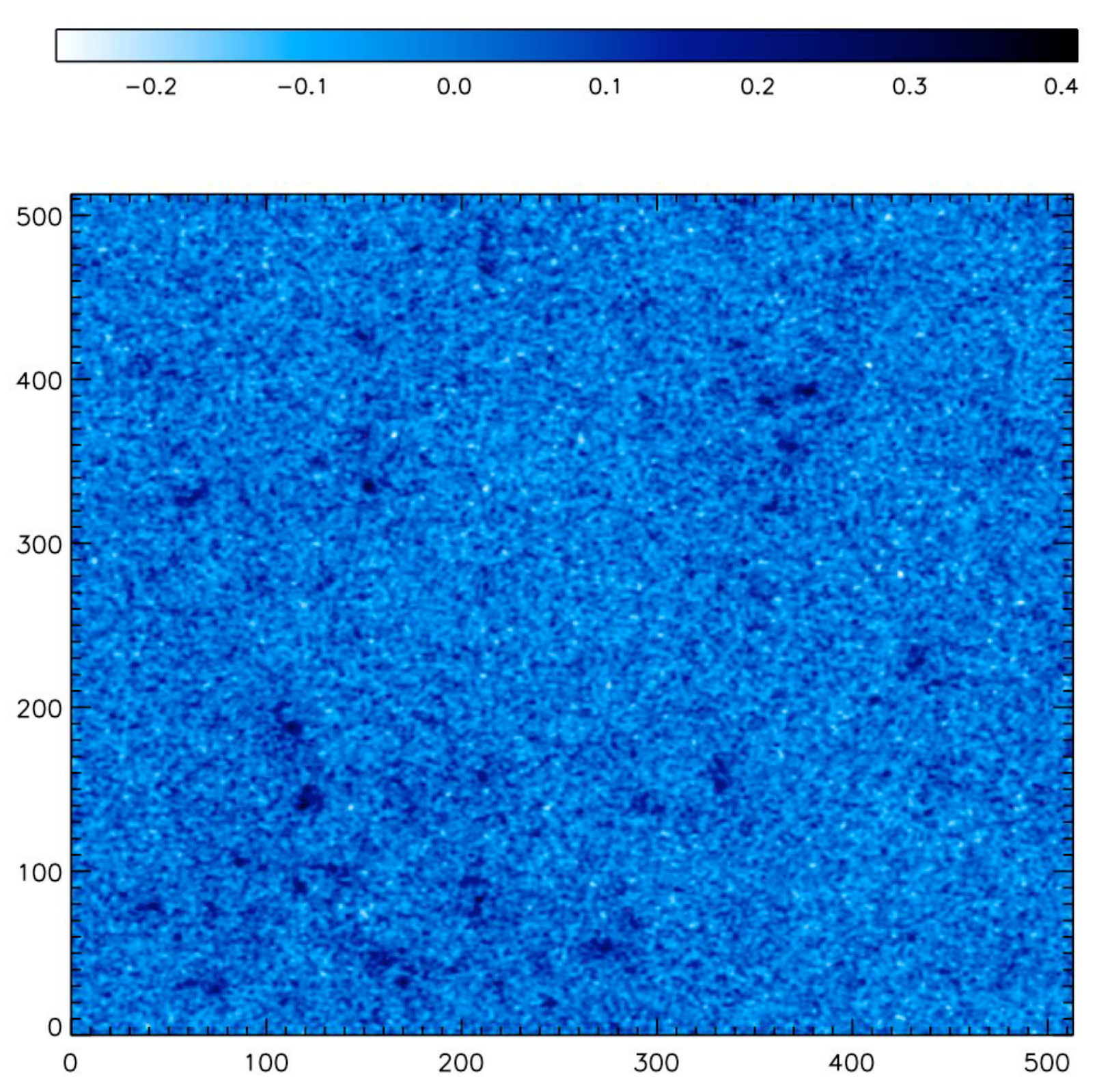}
\hspace{0.2cm}
\includegraphics[width=5.6cm, height=6.55cm]{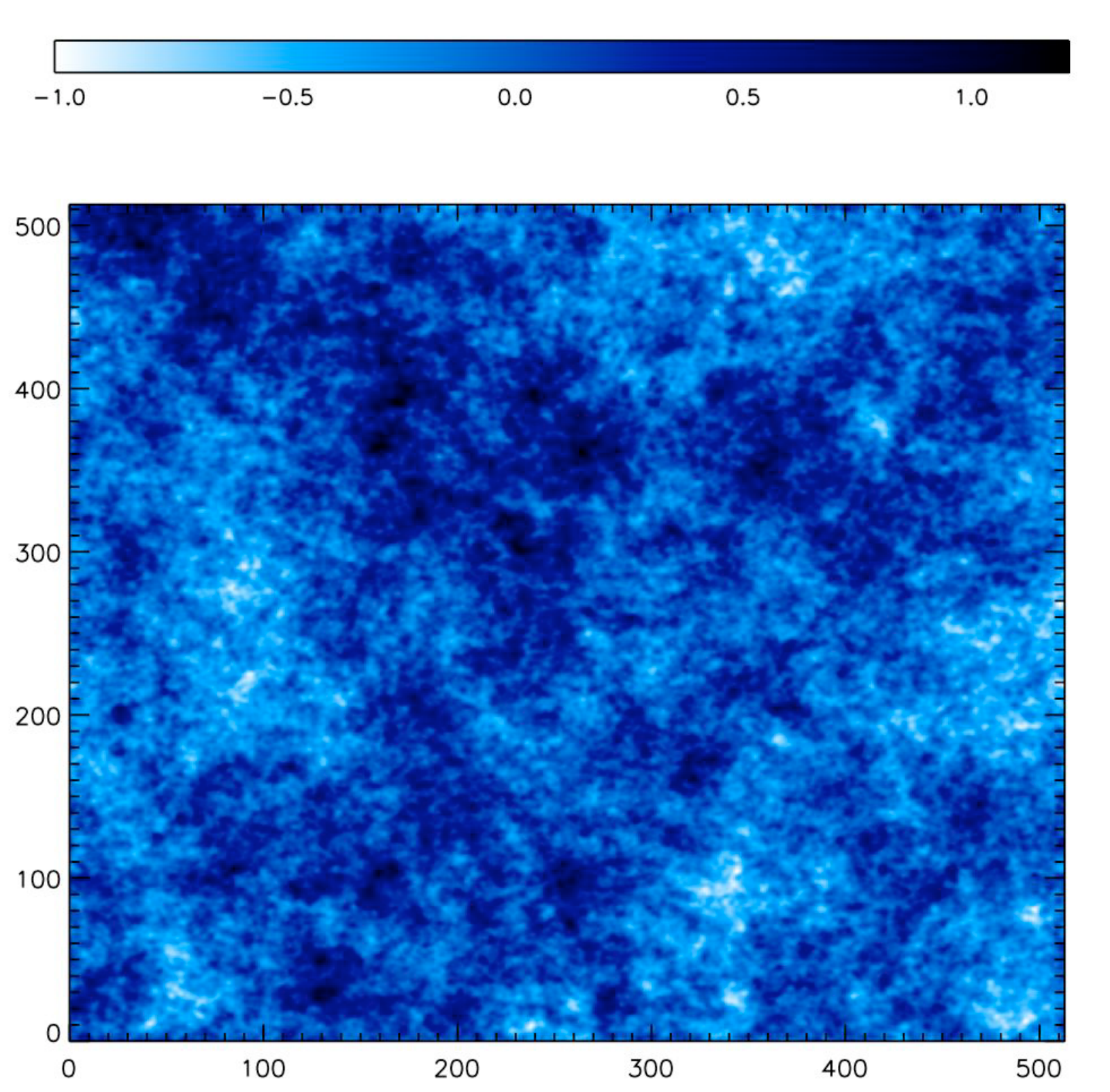}
}}
\caption{The 3 convergence maps have been smoothed with a Gaussian kernel of 15". {\bf Left} : Simulated convergence map $\kappa$ for a $\Lambda$CDM model. The field is $2^{\circ}$ x $2^{\circ}$ and is obtained from N-body simulations with 256$^3$ particles for a blocksize of $160h^{-1}$ Mpc. {\bf Middle:} Convergence map $\kappa_n$ reconstructed from noisy shear measurements corresponding to space-based observations ($\sigma^\gamma_\epsilon = 0.3$ and $n_g=50$ gal/arcmin$^{2}$). {\bf Right:} Convergence map $\kappa_n$ reconstructed from noisy flexion measurements corresponding to space-based observations ($\sigma^{\mathcal{F}}_\epsilon \simeq 0.04$ arcsec$^{-1}$ and  $n_g=50$ gal/arcmin$^{2}$).}
\label{convergence}
\end{figure*}

The left panel of Fig.~\ref{convergence} shows a simulated convergence map derived from ray-tracing through N-body cosmological simulations \citep{code:teyssier02}. The cosmological model is taken to be a concordance $\Lambda$CDM model with parameters $\Omega_M=0.3$, $\Omega_{\Lambda}=0.7$, $h=0.7$ and $\sigma_8 = 0.9$. The simulation contains $256^3$ particles with a box size of $160h^{-1}$ Mpc. The resulting convergence map covers 2 x 2 degrees with 512 x 512 pixels and assumes a galaxy redshift of 1. The typical standard deviation values of $\kappa$ are thus of the order of a few percent.

\subsubsection{Shear inversion problem}
The shear mass inversion problem consists of reconstructing the convergence field $\kappa({\mathbf \theta})$ from the measured shear field $\gamma_i({\mathbf \theta})$ by inverting equations (\ref{eq:gamma_psi}) and (\ref{eq:kappa_psi}). There are a number of approaches for doing this in the literature, and a comparison between the different local inversion methods has been carried out by \cite{inversion:seitz96}. 

To simplify the comparison with flexion, we will use the global shear inversion method that is presented in \cite{wlens:starck06} because a similar formalism exists for flexion.
For this purpose, we take the Fourier transform of the previous equations and
obtain 
\begin{equation}
\hat{\gamma_i} = \hat{P_i} \hat{\kappa},~~~i=1,2,
\label{gamma}
\end{equation}
where the hat symbol denotes Fourier transforms and
\begin{eqnarray}
\hat{P_1}(\mathbf k) & = & \frac{k_1^2 - k_2^2}{k_1^2+k_2^2}, \nonumber \\
\hat{P_2}(\mathbf k) & = & \frac{2 k_1 k_2}{k_1^2+k_2^2}.
\end{eqnarray}
The ideal shear maps $\gamma_i$ without noise can then be estimated from the convergence map $\kappa$. \\

By noting that $\hat{P_1}^2+\hat{P_2}^2=1$, an estimator of the mass distribution $\kappa$ can easily be derived by inversion 
\begin{equation}
\hat{\kappa}_n =  \hat{P_1} \hat{\gamma}_{1}+
  \hat{P_2}\hat{\gamma}_{2}.
\label{eqn_reckE}
\end{equation}

The deformation induced by weak gravitational lensing on a single galaxy is very weak compared to its intrinsic ellipticity. The lensing signal therefore must be extracted from the galaxy image ellipticity by assuming the intrinsic ellipticity is randomly oriented in the absence of gravitational lensing. The observed shear $\gamma_{i,n}$ is then obtained by averaging over a finite number of galaxies and, therefore, is noisy. 
The relationship between the observed data $\gamma_{1, n},\gamma_{2, n}$ binned in pixels
of area $A$ and the true convergence map $\kappa$ are given by:
\begin{equation}
\label{eq_gamma}
\gamma_{i,n} = P_i * \kappa + N_i^{\gamma},
\end{equation}
where $N_1^{\gamma}$ and $N_2^{\gamma}$ are white Gaussian noise with zero mean and standard deviation $\sigma_n \simeq \sigma^{\gamma}_{\epsilon}/\sqrt{N_g}$, where
$N_g = n_g A$ is the average number of galaxies in a pixel ($n_g$
is the average number of galaxies per area unit and A is the pixel area in the same unit). The rms shear
dispersion per galaxy $\sigma^{\gamma}_{\epsilon}$ arises both from measurement errors and the intrinsic shape dispersion of galaxies. In this analysis, we will assume $\sigma^{\gamma}_{\epsilon} \simeq 0.3$ as is approximately found for ground-based and space-based weak lensing
surveys  \citep{astro:brainerd96}. Typical values for the galaxy surface density for weak lensing are $n_g \sim 10$  gal/arcmin$^2$ for ground-based surveys and $n_g \sim 50$ gal/arcmin$^2$ for relatively deep space-based surveys.
In presence of noise, the estimator of the convergence $\kappa$ is:
\begin{equation}
\hat{\kappa}_n =  \hat{P_1} \hat{\gamma}_{1n}+
  \hat{P_2}\hat{\gamma}_{2n}.
\label{eqn_reckEE}
\end{equation}

\subsection{Flexion formalism}

\subsubsection{The third-order of the gravitational lensing potential}

As for shear, the flexion estimation can be calculated with shapelets \citep{flexion:goldberg05,flexion:bacon06,flexion:massey07} or by directly measuring the higher-order moments of the galaxy image \citep{flexion:okura07}. Flexion has two components, $\mathcal{F}$ and $\mathcal{G}$. A third-order inversion can be performed to recover the convergence field $\kappa(\theta)$ from the flexion field $\mathcal{F}$ or $\mathcal{G}$. It has been shown by \cite{flexion:okura07}  that measurements of the second component of flexion $\mathcal{G}$ is more noisy than the first component $\mathcal{F}$. In what follows we will only be interested in $\mathcal{F}$. 

The flexion $\mathcal{F}_i({\mathbf \theta})$ is derived from the second-order shape of galaxies at positions ${\mathbf\theta}$ in the image. The flexion field $\mathcal{F}_i({\mathbf \theta})$ can be written in terms of the lensing potential $\psi({\mathbf\theta})$ as:
\begin{eqnarray}
\label{eq:flexion1} 
\mathcal{F}_1 = \frac{1}{2}(\partial_1^3 + \partial_1 \partial_2^2)\Psi, \nonumber \\ 
\mathcal{F}_2 = \frac{1}{2}(\partial_2^3 + \partial_1^2 \partial_2)\Psi. 
\end{eqnarray}

\subsubsection{Flexion inversion problem}
The flexion mass inversion problem consists of reconstructing the convergence field $\kappa({\mathbf \theta})$ from the measured flexion field $\mathcal{F}_i({\mathbf \theta})$ by inverting equations (\ref{eq:kappa_psi}) and (\ref{eq:flexion1}). 
The Fourier transform of the relation $\mathcal{F}_i = \partial_i \kappa$ gives: 
\begin{eqnarray}
\hat{\mathcal{F}_1} = -i k_1 \hat{\kappa}(k), \nonumber \\ 
\hat{\mathcal{F}_2} = -i k_2 \hat{\kappa}(k). 
\label{eq:flex1}
\end{eqnarray}
Then, an estimator of the convergence $\kappa$ can be estimated: 
\begin{equation}
\hat{\kappa} = \frac{ik_1}{k_1^2+k_2^2} \hat{\mathcal{F}}_1(k) + \frac{ik_2}{k_1^2+k_2^2} \hat{\mathcal{F}}_2(k).
\label{eq:flex2}
\end{equation}

In the same way as for shear maps, a measurement error can be associated with the flexion maps. The relations between flexion measurements $\mathcal{F}_{1,n}$, $\mathcal{F}_{2,n}$ and the convergence map $\kappa$ are given by:
 \begin{equation}
\mathcal{F}_{i, n} =  \partial_i  \kappa + N_i^{\mathcal{F}},
\label{eq:flex6}
\end{equation}
where $N_1^{\mathcal{F}}$ and $N_2^{\mathcal{F}}$ are the noise contributions with a mean equal to zero and a rms equal to $\sigma_n =\sigma_\epsilon^{\mathcal{F}}/\sqrt{N_g}$. The flexion measurement error $\sigma_\epsilon^{\mathcal{F}}$ is between $\sigma^{\mathcal{F}}_\epsilon = 0.01$ arcsec$^{-1}$ (at $z=0$) and $\sigma^{\mathcal{F}}_\epsilon = 0.1$ arcsec$^{-1}$ (at $z=1$). We choose $\sigma^{\mathcal{F}}_\epsilon \simeq 0.04$ arcsec$^{-1}$ as in \cite{flexion:bacon06}. In our study, the distribution of the flexion measurements is assumed Gaussian although it is not the case in real data. However, this will not affect the conclusions.

In presence of noise the estimator of the convergence $\kappa$ is: 
\begin{equation}
\hat{\kappa}_n = \frac{ik_1}{k_1^2+k_2^2} \hat{\mathcal{F}}_{1,n}(k) + \frac{ik_2}{k_1^2+k_2^2} \hat{\mathcal{F}}_{2,n}(k).
\label{eq:flex3}
\end{equation}

\section{Comparison}
\label{comp_shear_flex}
Flexion should dominate over shear on small scales \citep{flexion:bacon06} since 
flexion effects are higher-order deformation of the gravitational potential. Small scale mass distributions should therefore be covered with higher fidelity with flexion. But what happens when measurement errors are added to the data ?

\subsection{Shear noise properties}
\label{shear_noise}
The intrinsic ellipticity and the measurement errors on the shear estimation of background galaxies result in an additive Gaussian noise on each shear component (see equation \ref{eq_gamma}). The standard dispersion on the shear measurement is $\sigma^{\gamma}_\epsilon \simeq 0.3$ \citep{astro:brainerd96}.
The noise on the convergence map $\kappa_n$ is an additive noise $N^{\gamma}$:
\begin{equation}
\hat{\kappa}_n = \hat{\kappa}  + \hat{N}^{\gamma}
\label{eq:shear2}
\end{equation}
where : 
\begin{equation}
\hat{N}^{\gamma}  = \hat{P_1} \hat{N_1}^{\gamma} + \hat{P_2} \hat{N_2}^{\gamma}.
\label{eq:shear3}
\end{equation}

The noise $\hat{N}$ in $\hat{\kappa}_n$ is still white, Gaussian and uncorrelated. The noise is not amplified by the inversion, but $\hat{\kappa}_n$ can be dominated by noise if $\hat{N}$ is large, which happens in practice. 

To simulate space observations, a realistic white Gaussian noise has been added to simulated shear maps. The reconstructed convergence map is dominated by a white gaussian noise ($\sigma_n=0.181$). 
The middle panel of Fig.~\ref{convergence} shows the reconstructed convergence map smoothed by a Gaussian kernel of 15". The smoothing is used to enable the detection of some clusters.

\subsection{Flexion noise properties}
\label{bruit_flexion}
The measurement errors on the flexion estimation of background galaxies result in an additive Gaussian noise on each flexion component $\mathcal{F}_i$ (see equation \ref{eq:flex6}). The dispersion on the flexion measurement that comes essentially from the flexion measurement errors is chosen to be $\sigma^{\mathcal{F}}_\epsilon \simeq 0.04$ arcsec$^{-1}$. 
The noise appears on the convergence map $\kappa_n$ as an additive noise $N$:
\begin{equation}
\hat{\kappa}_n = \hat{\kappa} + \hat{N}^{\mathcal{F}},
\label{eq:flex4}
\end{equation}
 where 
 \begin{equation}
 \hat{N}^{\mathcal{F}} = \frac{i k_1 \hat{N}_1^{\mathcal{F}}+ i k_2 \hat{N}_2^{\mathcal{F}}}{k_1^2+ k_2^2}.
 \label{eq:flex5}
\end{equation}
The flexion measurement errors then result in an additive colored Gaussian noise whose power is a function of $1/k$.
The right panel of the Fig.~\ref{convergence} shows a convergence map recovered from simulated flexion measurements ${\mathcal{F}_{i,n}}$. As expected, the convergence map appears contaminated by a colored Gaussian noise whose power is inversely proportional to the frequency $k$. No cluster is detected despite the fact that the map has been smoothed by a Gaussian kernel.\\

\subsection{Comparison between shear noise and flexion noise}
\begin{figure}[htp!]
 \centerline{
\includegraphics[width=7.5cm, height=7.cm]{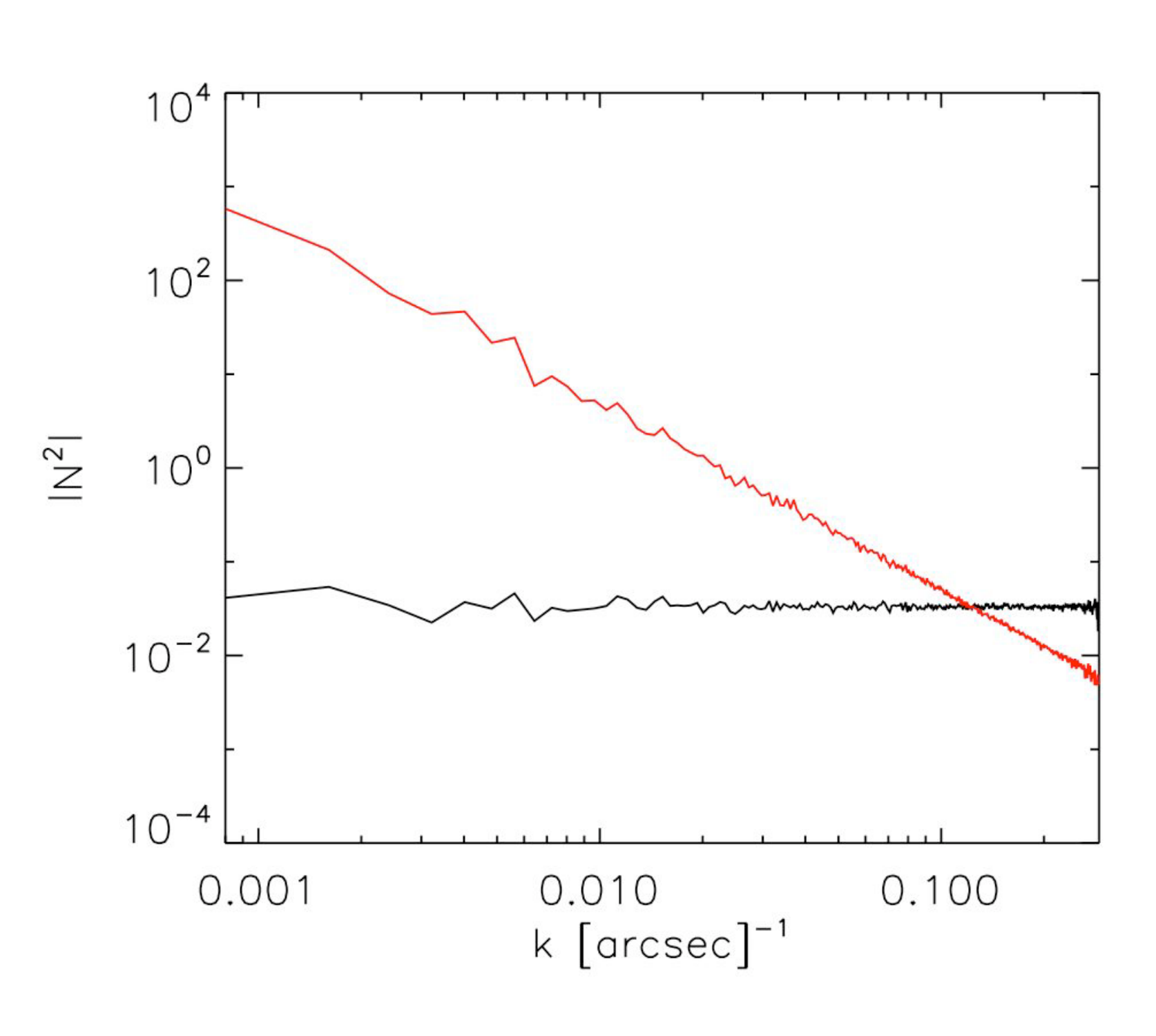}
}
\caption{Noise power spectrum on convergence map from shear measurements (in black) and from flexion measurements (in red) with realistic dispersions ($\sigma^\gamma_\epsilon = 0.3$ and $\sigma^\mathcal{F}_\epsilon = 0.04$ arcsec$^{-1}$) and assuming the galaxy density is the same for shear and flexion measurements ($n_g$ = 50 gal/arcmin$^2$).}
\label{crossing}
\end{figure}

In this paper, we compare the ability of flexion and shear to reconstruct the dark matter distribution. Since flexion dominates on small scales, we calculate here the scale at which flexion becomes dominant over shear. To do this, in Fig.~\ref{crossing}, we compare the noise power spectrum on convergence map obtained from shear measurements (solid black line) to the one obtained from flexion measurements (solid red line). The two solid lines have been obtained with realistic values of dispersion for space-based observations. The crossing of these two curves gives us the scale at which flexion becomes dominant over shear. As expected, the shear noise power spectrum is flat and the flexion noise power spectrum is inversely proportional to the frequency $k$.
 
The relations \ref{eq:shear3} and \ref{eq:flex5}  can be used to derive the analytic shear and flexion noise power spectrum:
\begin{eqnarray}
|\hat{N^\gamma}|^2 &\propto&\left(\frac{\sigma_\epsilon^\gamma}{\sqrt{N_g^\gamma}}\right)^2,  \rm{and} \nonumber \\
|\hat{N^\mathcal{F}}|^2 &\propto&\left(\frac{\sigma_\epsilon^\mathcal{F}}{\sqrt{N_g^\mathcal{F}} k}\right)^2.
\label{eq:flex8}
\end{eqnarray}
If the average number of galaxies in a pixel ($N_g$) is kept the same between shear and flexion measurements, the intersection of the two noise power spectra $k^T$ is given by $k^T=\frac{\sigma_\epsilon^\mathcal{F}}{\sigma_\epsilon^\gamma}$. If the standard values are used for shear and flexion dispersion ($\sigma^\gamma_\epsilon = 0.3$ and $\sigma^\mathcal{F}_\epsilon = 0.04$ arcsec$^{-1}$, $k^T=0.1333$ arcsec$^{-1}$) this corresponds to a scale of 7.5 arcsec. Thus, the flexion becomes interesting for scales smaller than 7.5 arcsec. To have at least a mean of 1 galaxy per pixel (with a pixel size of 7.5 arcsec), the galaxy density should be significantly larger than $n_g \sim 70$ gal/arcmin$^2$. Even if one has this galaxy density, all the pixels will not have a galaxy that falls inside, and one will have to deal with the problem of missing data.\\

Concerning the poor reconstruction of the convergence map from flexion (right panel of Fig.~\ref{convergence}), we note that attention has to be paid to the resolution of our simulation. The resolution of the simulation that we use for the study is 14 arcsec but many of the scales of interest for flexion are below this resolution. However, even with better resolution simulations these scales are not reachable when mapping with real data - especially if the missing data problem is not resolved.\\

\section{Discussion about published results on convergence reconstruction from flexion}
In the literature, several papers have tried to use the flexion to reconstruct the convergence map. Here is a discussion about the different studies. 
\subsection{Non-parametric convergence map reconstruction}
\begin{figure}[htp!]
\centerline{
\includegraphics[width=7.5cm, height=7.cm]{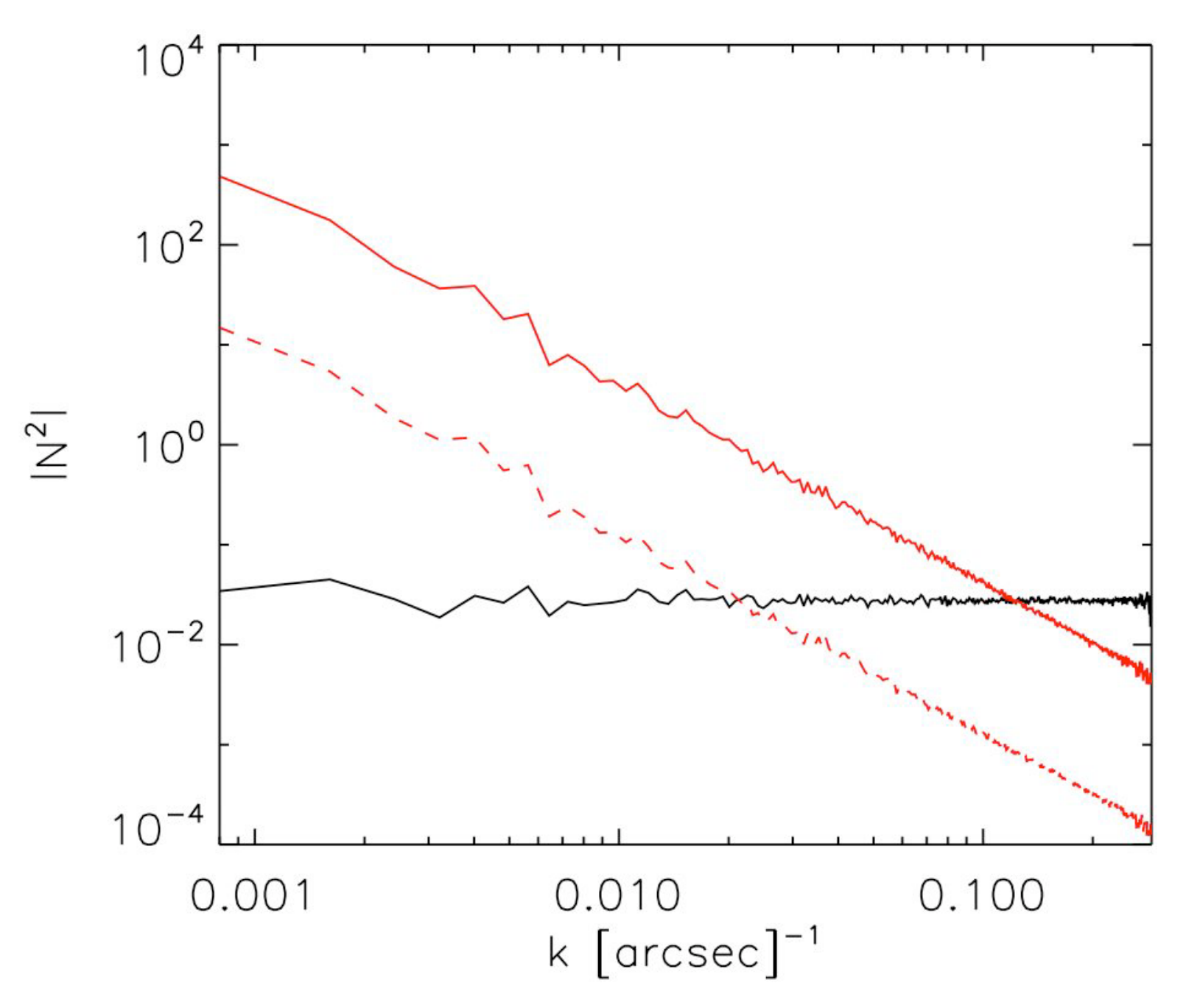}
}
\caption{Noise power spectrum on the convergence map obtained from realistic shear measurements corresponding to space-based observations (solid black line). The solid red line corresponds to the noise power spectrum on convergence map obtained from the realistic flexion measurements reported in \cite{flexion:bacon06} ($\sigma_e^{\mathcal{F}}$=0.04 arcsec$^{-1}$) and the dashed red line corresponds to the noise power spectrum obtained from the very optimistic flexion measurements that have been really used incorrectly in \cite{flexion:bacon06} ($\sigma_e^{\mathcal{F}}$=0.007 arcsec$^{-1}$).  We assume the galaxy density is the same for shear and flexion measurements and we adopt the optimistic galaxy density of \cite{flexion:bacon06} ($n_g$ = 60 gal/arcmin$^2$).}
\label{crossing_bacon}
\end{figure}

Reconstructions of convergence maps using flexion measurements were first introduced by \cite{flexion:bacon06}. In their paper, a (non-parametric) convergence map is reconstructed from simulated flexion measurements. The simulations have a galaxy density of $n_g=60$ gal/arcmin$^2$ and a reported flexion dispersion of $\sigma_e^{\mathcal{F}}$=0.04 arcsec$^{-1}$. However, an error in the reconstruction code meant that the true flexion dispersion was $\sigma_e^{\mathcal{F}}$=0.007 arcsec$^{-1}$, which would only be achievable for the highest signal-to-noise galaxies\footnote{Private Communication from Bacon, D.}.
Fig.~\ref{crossing_bacon} compares the noise power spectrum on the convergence map obtained from realistic shear measurements (solid black line) and from realistic flexion measurements (solid red line). The dashed red line gives the result from the optimistic flexion measurements used by \cite{flexion:bacon06} ($\sigma_e^{\mathcal{F}}$=0.007 arcsec$^{-1}$ and $n_g$ = 60 gal/arcmin$^2$). These values are optimistic since to achieve this dispersion the flexion of the highest signal-to-noise galaxies should be measured, which lead to a galaxy density significantly smaller than $n_g$ = 60 gal/arcmin$^2$. Doing so increases the ratio $\frac{\sigma_\epsilon^\mathcal{F}}{\sigma_\epsilon^\gamma}$, but decreases the ratio $\frac{n_g^\gamma}{n_g^\mathcal{F}}$ (see equation [\ref{eq:flex8}]) because of the small number of high signal-to-noise galaxies. At the end, the scale $k^T$ should remain almost the same. In \cite{flexion:bacon06}, the reconstruction fidelity from flexion measurements is therefore too optimistic and the result of the reconstruction should be closer to the right panel of Fig.~\ref{convergence}.

\begin{figure}[htp!]
\centerline{
\includegraphics[width=7.5cm, height=7.cm]{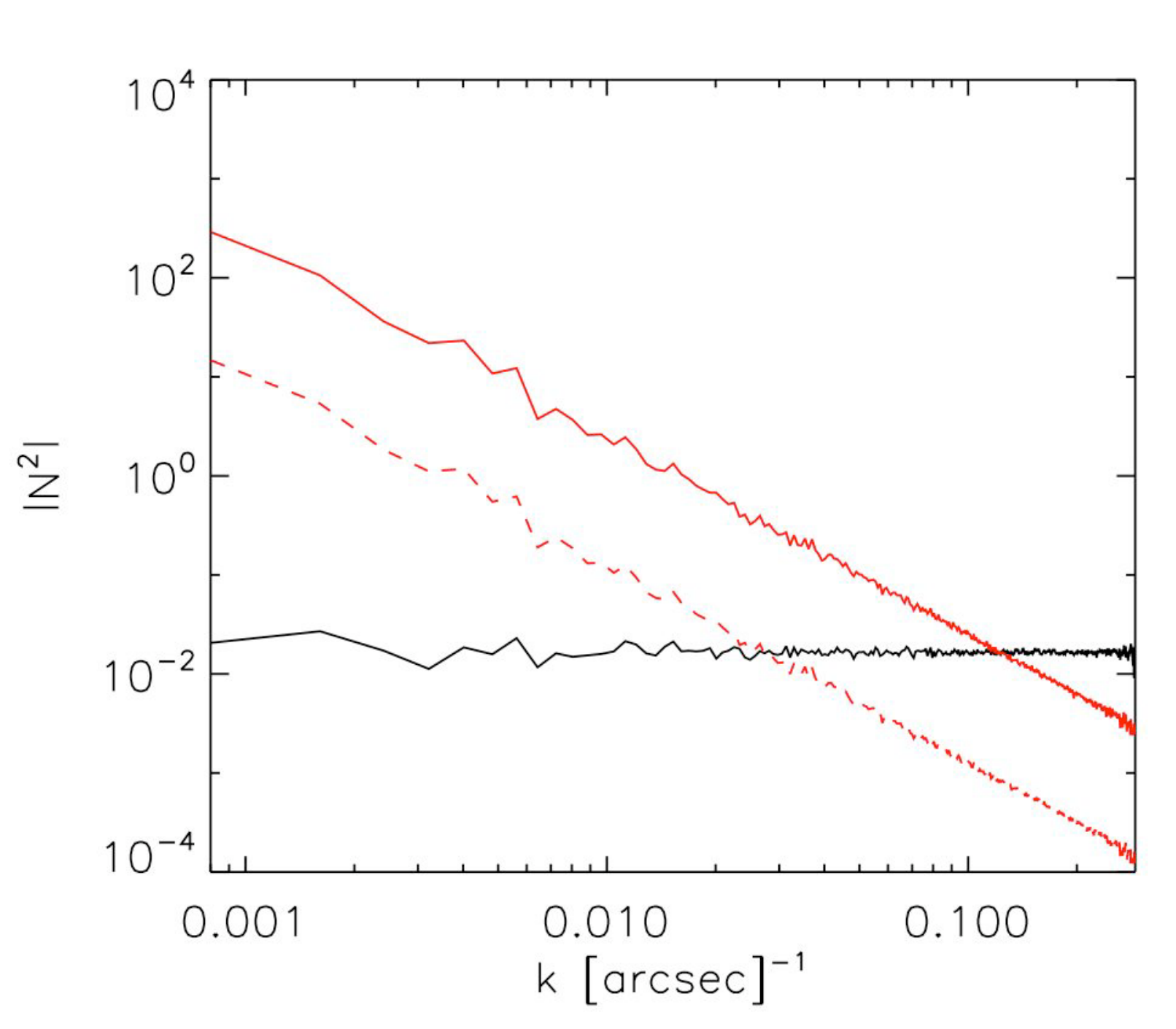}
}
\caption{Noise power spectrum on convergence map obtained from realistic shear measurements (solid black line) and from realistic flexion measurements (solid red line). The dashed red line corresponds to the noise power spectrum on convergence map obtained from the optimistic flexion dispersion ($\sigma_e^{\mathcal{F}}$=0.009 arcsec$^{-1}$) of Okura et al, 2007. We assume the galaxy density is the same for shear and flexion measurements and we adopt the optimistic galaxy density of Okura et al, 2007 ($n_g = 100$ gal/arcmin$^2$).}
\label{crossing_okura}
\end{figure}

In \cite{flexion:okura07}, the authors also use a (non-parametric) convergence map reconstruction from simulated flexion measurements. But the data are simulated with a rather optimistic galaxy density $n_g = 100$ gal/arcmin$^2$, and a very optimistic value has been chosen for the flexion measurement error $\sigma^\mathcal{F}_\epsilon=0.009$ arcsec$^{-1}$. The dashed red line of Fig.~\ref{crossing_okura} shows the noise power spectrum that should be obtained with this optimistic value of flexion dispersion. The intersection with the solid black line gives the scale below which the flexion dominates ($k^T$ = 50 arcsec). As shown in \cite{flexion:okura07}, the reconstruction of a binned convergence map by combining shear and flexion measurements is interesting with this optimistic flexion dispersion because the flexion is dominating for scales smaller than 50 arcsec. But, using real data with a realistic flexion dispersion, the result of the reconstruction of a binned convergence map from flexion measurements should be close to what is shown in the right panel of Fig.~\ref{convergence}. 

In \cite{flexion:okura08}, this convergence map reconstruction method is applied to real data (ground-based Subaru data). The flexion dispersion is found to be $\sigma_e^{\mathcal{F}}=0.11245$ arcsec$^{-1}$. The galaxy density is very small at $n_g = 7.75$ gal/arcmin$^2$ (only the 791 brightest galaxies have been selected) and the field is 9' x 9' sampled with a grid of 256 x 256 pixels. Its 791 galaxies to divide into 65536 pixels, which means that only 1\% of pixels have a galaxy inside. In the paper, there is no mention about the convergence map reconstruction problem from incomplete shear maps given that 99 \% of data are missing (see \cite{gaps:pires09} for more details about the missing data problem). No detections should be possible in convergence maps obtained from these flexion measurements. But, it is difficult to characterize the noise properties of the convergence maps produced by this method and then to access to the significance of the detections.

In \cite{flexion:leonard09}, an aperture mass is used to reconstruct the convergence map from simulated flexion measurements. The flexion dispersion is taken $\sigma_e^{\mathcal{F}}=0.1$ arcsec$^{-1}$ and the galaxy density $n_g = 35$ gal/arcmin$^2$. The authors claim to reconstruct substructures, but except for the peak that is detected with more than a 3$\sigma$ detection level, no substructure is detected with more than a 2$\sigma$ detection level. It should also be noted that a fair comparison with the aperture mass for shear is not carried out in this paper. By consequence, no conclusions about the utility of flexion measurements to reconstruct substructures can be draw from this study. However, the authors are working on this and a paper will be submitted later this year.\footnote{Private communication from Leonard, A.}

\subsection{Parametric convergence map reconstruction}
In \cite{flexion:leonard07}, a parametric convergence reconstruction is performed to reconstruct A1689 cluster (from HST ACS space-based data), which is one of the biggest and most massive known galaxy clusters. The galaxy density is important ($n_g = 75$ gal/arcmin$^2$) because of the magnification effect. The measurements were carried out on stacked images, which resulted in better shape measurement accuracy ($\sigma_e^{\mathcal{F}}=0.029$ arcsec$^{-1}$). In this study, the galaxy-galaxy flexion signal has been used to show that foreground galaxies are well-fitted by a singular isothermal sphere with a characteristic dispersion $\sigma_v$. Then, for each confirmed foreground galaxy, the dispersion $\sigma_{v,i}$ is estimated from their flexion effect on background galaxies. Therefore, the mass reconstruction is modeled as the sum of the fits obtained for each foreground galaxy. This method is rather reliable because it depends on the visible distribution of the cluster. Then, it offers a way to include the flexion measurements in the reconstruction method. However, the measure of the dispersion $\sigma_{v,i}$ for each foreground galaxy remains very noisy and the reconstruction takes no account for the possible presence of dark haloes in the cluster.

\section{Conclusion}
The aim of this paper is to compare the ability of shear and flexion to reconstruct convergence maps. A comparison between shear and flexion, taking into account the noise contributions, has been carried out. Using noise simulations, we have shown that flexion becomes more interesting than shear on scales smaller than the scale containing one galaxy (pixel scale). Consequently, the flexion measurements should not be used alone to reconstruct a binned convergence map because the flexion is dominating on scales beyond the pixel scale. The literature contains several papers that try to use flexion to reconstruct convergence map but, their results are not convincing. 

Nonetheless, flexion has already been detected and can still be used to measure the statistical properties of substructures in dark matter halos on very small scales \citep{flexion:bacon09}. 

Concerning convergence map reconstruction, it is now clear that flexion should not be used alone. However, it does help to add the flexion of galaxies in mass reconstruction from shear measurements. The question is, how should shear and flexion be combined for optimal results ? 
In \cite{flexion:leonard07,flexion:shapiro10}, the authors propose a way of doing this but moving beyond this, more working is still needed to find a Bayesian reconstruction method for the inclusion of flexion.

\end{document}